# Study of Cullum's and Willoughby's Lanczos method for Wilson fermions ⋆

Thomas Kalkreuter [1]

*Institut für Physik, Humboldt-Universität, Invalidenstraße 110, D-10099 Berlin, Germany*

The Lanczos method of Cullum and Willoughby is studied for euclidean Wilson fermions in quenched and unquenched SU(2) gauge fields on lattices of volume ranging from $4^4$ to $16^4$. The method is reliable even on larger lattices, but its cost for the computation of a given fraction of the spectrum grows (approximately) with the square of the lattice volume. We investigate the convergence behaviour and show that it is closely linked with the local spectral density. Complete spectra are determined on lattices up to $8^3 \cdot 12$. For configurations where all eigenvalues are computed, we give numerical values for the fermionic determinants and results for spectral densities. Determinants are also given for staggered fermions whose quenched and unquenched spectra were studied in a previous publication.

⋆ Work supported by Deutsche Forschungsgemeinschaft, grant Wo 389/3-1.
[1] Electronic address: kalkreut@linde.physik.hu-berlin.de

## 1 Introduction

In order to study questions of chiral symmetry breaking [1–6] and universality [7,8], and also in the context of Lüscher's fermion algorithm [9–11], one is interested in the eigenvalues of the gauge covariant Dirac operator (or $\gamma_5$ times the Dirac operator) which are close to the origin. One candidate method which comes to mind to determine eigenvalues numerically is a Lanczos algorithm [12]. Variants of this method have been used in lattice field theory for a long time, see e.g. Refs. [13,2] for staggered fermions and Refs. [14,3] for Wilson fermions.

Presently there is renewed interest in the Lanczos method, for instance in Refs. [11,6], but the topics of these papers are such that the algorithmic features and difficulties connected with a Lanczos procedure were not addressed. The present paper complements this point. We perform a further algorithmic investigation of the Lanczos method in the context of lattice gauge theory which extends previous studies. We are also interested in a comparison of the partially converged Lanczos method (i.e. the case that just eigenvalues close to the origin are to be computed) with an accelerated conjugate gradient algorithm for low-lying eigenvalues [15,16].

The focus of the present paper will be on Cullum's and Willoughby's Lanczos method [17] applied to euclidean Wilson fermions in quenched and unquenched SU(2) gauge fields on lattices of volume ranging from $4^4$ to $16^4$. The operators under consideration are

$$Q = \gamma_5 \, (D + m) \, / \, (8 + m) \tag{1}$$

and its square. $(D + m)$ is the massive Dirac operator for Wilson fermions,

$$\begin{aligned} &[(D+m)\psi](x) = \\ &\frac{1}{2\kappa}\,\psi(x) - \frac{1}{2}\sum_{\mu=1}^{4} \quad \{\, (\mathbb{1} - \gamma_\mu)\, U(x, x+\mu)\, \psi(x+\mu) \\ &\qquad\qquad\qquad\qquad + (\mathbb{1} + \gamma_\mu)\, U(x, x-\mu)\, \psi(x-\mu)\} \ . \end{aligned} \tag{2}$$

Here $\kappa = (2m + 8)^{-1}$ denotes the hopping parameter, $x \pm \mu$ is the nearest neighbour site of $x$ in $\pm\mu$-direction, and $U(x, x \pm \mu)$ is the gauge field on the link $(x, x \pm \mu)$. In the present paper we impose periodic boundary conditions. The operator $Q$ is hermitean, and it is normalized such that its eigenvalues are between -1 and 1.

In some situations one does not have to know the sign of the eigenvalues of $Q$ but only their absolute magnitude. This means that one can equally well determine (a part of) the spectrum of $Q^2$. For example, one can probe the chiral limit by means of the "pion norm" which depends on the eigenvalues of $Q^2$ as discussed in Ref. [1]; or in Lüscher's fermion algorithm [9,10] for an even number of flavours the small eigenvalues of $Q^2$ can be used to correct for possible systematic errors in case that the polynomial approximation to the function $1/s$ is too poor at the lower end of the spectrum. If one is in such a situation, one can run the Lanczos procedure either with $Q$ or with $Q^2$. We show that it is advantageous to diagonalize the unsquared operator which means that one could speed up computations as in Ref. [11].

Further points which will be discussed are given in the following overview. We concentrate on algorithmic aspects, physical questions will be addressed in a subsequent paper. In Sec. 2 we start by recalling Cullum's and Willoughby's Lanczos method. This method is reliable even on larger lattices, but its cost for the computation of a given fraction of the spectrum grows (approximately) with the square of the lattice volume. Issues of the convergence behaviour of the partially converged Lanczos method on lattices up to $16^4$ are described in Sec. 3. They reveal again the well-known fact that there is no black-box Lanczos routine, but one needs experience with the operator under consideration. In Sec. 4 we turn to the computation of complete spectra. It is shown that the convergence behaviour is closely related with the local spectral density. Complete spectra are determined on lattices up to $8^3 \cdot 12$ (and almost on a $12^4$ lattice where 165884 of the 165888 eigenvalues were found). For configurations where all eigenvalues are computed, we give results for the fermionic determinants in Sec. 5. In this section we also quote values of determinants of quenched and unquenched staggered fermions whose spectra were determined in a previous publication [18]. Finally, we present spectral densities in Sec. 6, and we end with some conclusions and a comparison with the algorithm of Ref. [16].

## 2 Cullum's and Willoughby's Lanczos method

The Lanczos procedure is a technique that can be used to solve large, sparse, symmetric or hermitean eigenproblems [2] [12]. The idea is to transform a given hermitean $n \times n$ matrix $A$ into a similar symmetric tridiagonal matrix $T = V^{-1}AV$ with unitary $V$, and then $T$ is diagonalized. The transformation of $A$ can be performed iteratively. If one writes $V = (v_1, v_2, \ldots, v_n)$ with column

[2] In Refs. [13,14,3] the authors used a non-hermitean Lanczos method and come to the conclusion that it works on small lattices.

vectors $v_i$ ("Lanczos vectors") and

$$T = \begin{pmatrix} \alpha_1 & \beta_1 & & \\ \beta_1 & \alpha_2 & \ddots & \\ & \ddots & \ddots & \beta_{n-1} \\ & & \beta_{n-1} & \alpha_n \end{pmatrix} \quad , \tag{3}$$

then $AV = VT$ is equivalent to

$$\begin{aligned} Av_1 &= \alpha_1 v_1 + \beta_1 v_2 \ , \\ Av_i &= \beta_{i-1} v_{i-1} + \alpha_i v_i + \beta_i v_{i+1} \quad \text{for } i = 2, \ldots, n-1, \\ Av_n &= \beta_{n-1} v_{n-1} + \alpha_n v_n \ . \end{aligned} \tag{4}$$

Given an initial — generally random — vector $v_1$ and using the orthonormality among the $v_i$ with respect to the canonical scalar product $\langle \cdot , \cdot \rangle$, one can determine iteratively from these equations: $\alpha_i = \langle v_i, Av_i \rangle$, $\beta_i^2 = (\langle v_i, A^2 v_i \rangle - \alpha_i^2 - \beta_{i-1}^2)$, (with $\beta_0 \equiv 0$), and $v_{i+1} = \beta_i^{-1}(Av_i - \beta_{i-1} v_{i-1} - \alpha_i v_i)$ as long as $\beta_i \neq 0$ (otherwise the iteration is stopped). Note that $T$ will be a real matrix also in case that $A$ is complex. The sign of $\beta_i$ is arbitrary.

In exact arithmetic, the Lanczos iteration should finish after at most $n$ steps and the last equation in (4) would be automatically fulfilled. For this case there exists a convergence theory for which we refer to the literature [12]. In practice, however, there are severe problems with a straightforward implementation of the Lanczos procedure [12,17]. These problems are caused by rounding errors and loss of orthogonality among the Lanczos vectors. [3] In principle the latter problem can be circumvented by storing all the Lanczos vectors and enforcing orthogonality among them by hand. However, then one is restricted to small lattices because of computer memory or I/O limitations.

In Cullum's and Willoughby's proposal [17] one performs no reorthogonalization, and one continues the iteration (4) for an a priori unspecified count. In this way a sequence of $j \times j$ tridiagonal matrices $T^{(j)}$, $j = 1, 2, \ldots$ is generated. The diagonal elements of $T^{(j)}$ are $\alpha_i$, $i = 1, \ldots, j$, and the off-diagonal entries are $\beta_i$, $i = 1, \ldots, j-1$. As a technical point we note that the $\alpha_i$, $\beta_i$ and $v_{i+1}$ will not be computed according to the formulas given above, but rather there exists a particular form of the recursion which has proven most stable, see [17] and [12, Algorithm 9.2.1 and remark on p. 492].

[3] This loss of orthogonality is not necessarily due to the accumulation of round-off errors [12].

The common belief is that generally the extremal eigenvalues of $T^{(j)}$ with increasing $j$ are progressively better estimates of the extremal eigenvalues of $A$. Eventually, all the different eigenvalues of $A$ will appear as eigenvalues of some $T^{(j)}$ with $j$ generally larger than $n$. However, because of rounding errors and loss of orthogonality among the Lanczos vectors, there will also appear so-called "spurious" eigenvalues. These are eigenvalues of $T^{(j)}$ but not (approximate) eigenvalues of $A$.

For the solution to the problem of identifying spurious eigenvalues and coping with their presence, Cullum and Willoughby give the following recipe [17]. One compares the eigenvalues of $T^{(j)}$ with the eigenvalues of a matrix $T_2^{(j)}$ which equals $T^{(j)}$ with the first row and first column deleted. If a simple eigenvalue of $T^{(j)}$ is also an eigenvalue of $T_2^{(j)}$, then this eigenvalue is spurious. (We remark that the authors of Refs. [13,14,2] work with $T^{(j-1)}$ instead of $T_2^{(j)}$. They identify approximate eigenvalues of $A$ by having the property to be eigenvalues of both $T^{(j-1)}$ and $T^{(j)}$ whereas spurious eigenvalues are different.)

A problem which remains in the Cullum-Willoughby algorithm is that the multiplicities of the eigenvalues of $T^{(j)}$ do not reflect the correct multiplicities of the eigenvalues of $A$. However, this is neither a problem for the present study as we shall see, nor was it a problem in a previous study for staggered fermions [18].

## 3 Convergence behaviour of the Lanczos method for eigenvalues close to and furthest from the origin

Henceforth the term Lanczos method refers to Cullum's and Willoughby's variant. The tridiagonal matrices $T^{(j)}$ and $T_2^{(j)}$ were diagonalized by means of the Numerical Recipes routine "tqli" [19] which implements the QL algorithm with implicit shifts. Two eigenvalues were counted as different when they differed by more than $10^{-10}$. This number is arbitrary but it is chosen such that it is small compared with the gaps in the spectra, and large compared with round-off errors. The computer program was checked for gauge covariance, and it was also verified that free spectra are obtained correctly, except for multiplicities.

We turn first to the convergence behaviour of the Lanczos method for the operator $Q$ defined in (1), and for eigenvalues $\lambda$ close to and furthest from the origin. The eigenvalues of $T^{(j)}$ were monitored as a function of $j$. Some results in individual gauge field configurations [4] are collected in Table 1. The entry

[4] The numerical data presented in this section were obtained from points in the $\beta$-$\kappa$-plane which are not in the immediate neighbourhood of a critical point.

Table 1
Number of different eigenvalues (EVs) found after $j$ iterations of the Lanczos procedure for the operator $Q$. The examples are taken at $\kappa = 0.15$, and $\beta$ denotes the coupling constant of the SU(2) gauge field part of the action.

| | $12^4$, $\beta = 2.4$, quenched | | $16^4$, $\beta = 2.12$, unquenched | | $16^4$, $\beta = 2.3$, unquenched | |
|---|---|---|---|---|---|---|
| $j$ | EVs $T^{(j)}$ | "good" EVs | EVs $T^{(j)}$ | "good" EVs | EVs $T^{(j)}$ | "good" EVs |
| 500 | 500 | 500 | 500 | 500 | 500 | 500 |
| 1000 | 992 | 988 | 1000 | 998 | 1000 | 998 |
| 2000 | 1945 | 1929 | 1987 | 1981 | 1982 | 1970 |
| 3000 | 2874 | 2842 | 2962 | 2947 | 2933 | 2920 |
| 4000 | 3777 | 3730 | 3921 | 3898 | 3893 | 3840 |
| 5000 | 4684 | 4599 | 4882 | 4839 | 4825 | 4749 |
| 6000 | 5561 | 5453 | 5814 | 5771 | 5707 | 5660 |
| 7000 | 6426 | 6293 | 6759 | 6692 | 6632 | 6563 |
| 8000 | 7271 | 7119 | 7693 | 7608 | 7552 | 7460 |
| 9000 | 8115 | 7937 | 8620 | 8518 | 8440 | 8347 |
| 10000 | 8966 | 8746 | 9541 | 9420 | 9334 | 9229 |

"EVs $T^{(j)}$" gives the total number of different eigenvalues of $T^{(j)}$, and "good EVs" means the number of different eigenvalues which are not spurious according to Cullum's and Willoughby's criterion. Note, however, that a "good" eigenvalue is generally only an approximation to an eigenvalue of $Q$, but this approximation has not necessarily converged.

A "good" eigenvalue turns into an "exact" one after convergence. Fig. 1 shows the convergence of the lowest positive eigenvalues and of the highest ones on an unquenched $16^4$ lattice at $\beta = 2.3$, $\kappa = 0.15$. There exists an almost point reflection symmetry in that the picture for the highest negative and for the lowest eigenvalues looks practically the same if one changes signs at the axes. On other lattices the figures look qualitatively similar.

In the runs where complete spectra were determined (see Sec. 4) the following observation could be confirmed [17]: Accurate approximations to eigenvalues of $Q$ will be stabilized eigenvalues of $T^{(j)}$. This means that if an eigenvalue has converged for some $j$, then it will be an "exact" eigenvalue of $T^{(j)}$ for any larger value of $j$. Hence, we infer from Fig. 1 that not only the extremal eigenvalues converge fastest but also the eigenvalues close to zero. This fact was already pointed out in Refs. [13,14]. The reason is that the convergence is fastest in those regions of the spectrum where the spectral density is lowest, and in the configuration of Fig. 1 this is the case close to and furthest from the origin. We will make the connection between convergence properties and the local spectral density more precise in Sec. 4.

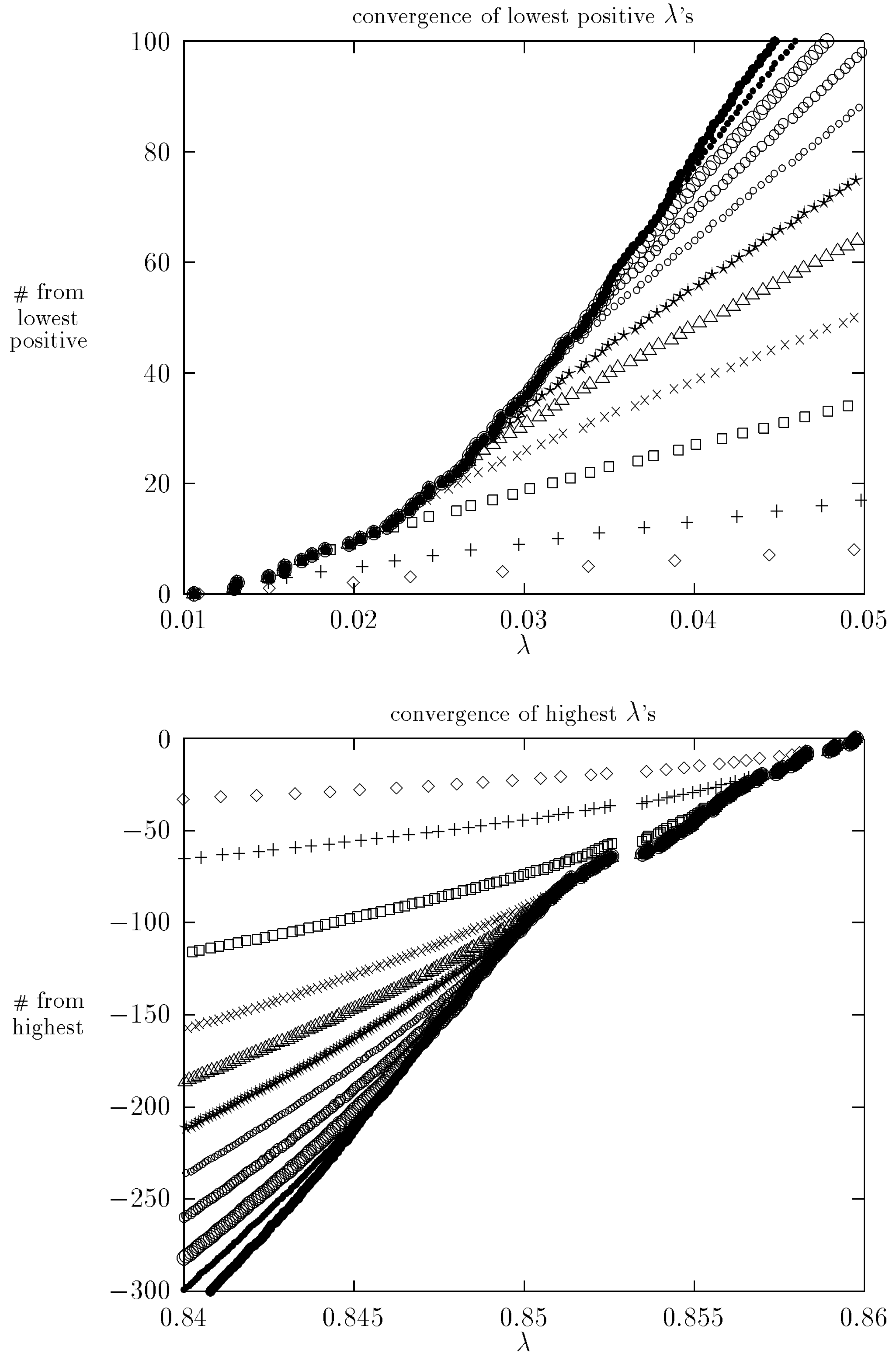

Fig. 1. Convergence of the lowest positive eigenvalues (top) and of the highest ones (bottom) of the unquenched $Q$ as found by the Lanczos method on a $16^4$ lattice, $\beta = 2.3$, $\kappa = 0.15$. From "$\diamond$" to "$\bullet$" the curves show the "good" eigenvalues of $T^{(j)}$ found with $j$ = 500, 1 000, 2 000, 3 000, 4 000, 5 000, 6 000, 7 000, 8 000, 9 000, and 10 000 iterations, respectively.

In the regions of lowest spectral density the eigenvalues usually converge in an "ordered" fashion by which we mean the following. Between any two converged eigenvalues there is no "good" eigenvalue which has not yet converged also or which will not converge within a few more iterations. This observation is in accordance with Ref. [2] for staggered fermions.

After an eigenvalue has converged, it starts to replicate [17] which means that it appears as a multiple eigenvalue of $T^{(j)}$, even if it is only a simple eigenvalue of $Q$. The converse can also happen: A truly degenerate eigenvalue may only appear as a simple eigenvalue of $T^{(j)}$ for some $j$. In any case, the multiplicities of the eigenvalues of $T^{(j)}$ do not reflect the multiplicities of the eigenvalues of $Q$.

As mentioned in the introduction we are interested in a comparison of the partially converged Lanczos method with an accelerated conjugate gradient algorithm for low-lying eigenvalues [15,16]. This practical conjugate gradient variant determines low-lying eigenvalues of $Q^2$ with rigorous practical error bounds. Analogous error bounds do not exist for the Lanczos method, [5] but we are able to monitor the accuracy of the non-spurious eigenvalues of $T^{(j)}$ a posteriori by referring to the converged (stabilized) values. We squared and sorted the "good" eigenvalues of $T^{(j)}$ and performed this convergence monitoring. By requiring a relative accuracy of $10^{-4}$, we obtain the results of Table 2. These results can be compared with Table 1 of Ref. [16], yielding an inferiority of the accelerated conjugate gradient method by a factor of $5-8$ when one counts only the number of $Q\times$vector multiplications which are necessary for a relative error of $10^{-4}$. The iteration number $j$ where this accuracy is reached is not known immediately, but we note that when the relative error is already $10^{-4}$, then it will be reduced further very quickly. Hence, the entries in Table 2 will not change much when we use the stabilization of the eigenvalues as an indication for their convergence.

Instead of running the Lanczos procedure with $A = Q$ one can alternatively use it with $A = Q^2$ if one is not interested in the signs of the eigenvalues. Doing so, the results for the test configurations presented in Tables 1 and 2 and in Fig. 1 remain similar. However, now one Lanczos iteration involves two instead of one application of $Q$, and the cost is roughly doubled. Therefore, it is advantageous to diagonalize the unsquared operator in case that one requires the lowest or highest eigenvalues of $Q^2$. This statement will become more and more significant, and the cost when working with $Q^2$ will further grow, as a critical point is approached in the $\beta$-$\kappa$-plane. This discussion is resumed in Sec. 6.

[5] Cullum and Willoughby quote error estimates [17], but they are not practical in lattice gauge theory, because either all eigenvalues of both $T^{(j)}$ and $T_2^{(j)}$ are required or one needs eigenvector approximations.

Table 2
Examples for the numbers of lowest eigenvalues of $Q^2$ with a relative error $< 10^{-4}$, found after $j$ iterations of the Lanczos procedure for the operator $Q$. The lowest eigenvalue of $Q^2$ is denoted by $\lambda_1^2$.

| j | $6^4$ lattice, $\kappa = 0.15$, $\beta = 1.80$, quenched, $\lambda_1^2 = 5.240 \cdot 10^{-3}$ | $6^3 \cdot 12$ lattice, $\kappa = 0.15$, $\beta = 2.12$, unquenched, $\lambda_1^2 = 1.332 \cdot 10^{-3}$ | $8^3 \cdot 12$ lattice, $\kappa = 0.15$, $\beta = 2.12$, unquenched, $\lambda_1^2 = 8.098 \cdot 10^{-4}$ | $8^4$ lattice, $\kappa = 0.20$, $\beta = 0.00$, quenched, $\lambda_1^2 = 1.592 \cdot 10^{-3}$ |
|---|---|---|---|---|
| 250 | 0 | 0 | 0 | 0 |
| 500 | 7 | 4 | 1 | 0 |
| 1000 | 22 | 21 | 9 | 2 |
| 2000 | 73 | 56 | 37 | 14 |
| 3000 | 111 | 85 | 67 | 28 |
| 4000 | 179 | 145 | 94 | 38 |
| 5000 | 254 | 182 | 119 | 50 |

| j | $12^4$ lattice, $\kappa = 0.15$, $\beta = 1.80$, quenched, $\lambda_1^2 = 4.752 \cdot 10^{-3}$ | $12^4$ lattice, $\kappa = 0.15$, $\beta = 2.40$, quenched, $\lambda_1^2 = 1.364 \cdot 10^{-4}$ | $16^4$ lattice, $\kappa = 0.15$, $\beta = 2.12$, unquenched, $\lambda_1^2 = 7.703 \cdot 10^{-4}$ | $16^4$ lattice, $\kappa = 0.15$, $\beta = 2.30$, unquenched, $\lambda_1^2 = 1.367 \cdot 10^{-4}$ |
|---|---|---|---|---|
| 500 | 0 | 0 | 0 | 0 |
| 1000 | 3 | 4 | 2 | 2 |
| 2000 | 16 | 18 | 8 | 8 |
| 3000 | 25 | 41 | 17 | 25 |
| 4000 | 37 | 62 | 23 | 38 |
| 5000 | 49 | 71 | 30 | 45 |
| 6000 | 57 | 99 | 49 | 59 |
| 7000 | 70 | 112 | 63 | 77 |
| 8000 | 79 | 135 | 75 | 101 |
| 9000 | 101 | 161 | 83 | 111 |

As can be seen from Table 2, the number of converged eigenvalues is not a simple function of $j$. In particular there is no linear relationship between the number of converged eigenvalues and $j$, so that a proceeding as in Ref. [11] might require some refinement. A comparison of Tables 1 and 2 also yields no simple connection between the number of "good" eigenvalues and the number of converged eigenvalues. And finally, one can conclude from Table 2 that the computational cost for a certain fraction of the total number of eigenvalues cannot be expected to increase with less than the square of the lattice volume.

## 4 Computation of complete spectra

Although usually one does not want to compute all eigenvalues in production runs, it is nevertheless worthwhile to consider also complete spectra in the present technical study. One reason is that one can derive valuable information about the quality of the eigenvalues which are obtained for $j \ll n$.

Let us start by looking at the convergence behavior of the Lanczos algorithm not only for the fastest converging eigenvalues but over the entire spectrum. We would like to concretize what is meant by the general statement of Ref. [17] that the local gap structure plays a role. An example with $n = 20\,736$ eigenvalues in an unquenched configuration on a $6^3 \cdot 12$ lattice at $\beta = 2.12$, $\kappa = 0.15$ is shown in Fig. 2. The Lanczos method was run both with $A = Q$ and with $A = Q^2$. Every 500 iterations the matrices $T^{(j)}$ and $T_2^{(j)}$ were diagonalized, and the "good" eigenvalues of $T^{(j)}$ were compared with reference data from $T^{(82944)}$ (see below for the choice of $j = 82\,944 = 32 \times 6^3 \cdot 12$). Every dot in Fig. 2 indicates a point where an eigenvalue $\lambda$ of $Q$ or $\lambda^2$ of $Q^2$ has converged. Diamonds show the appropriately scaled spectral densities $\rho$ (see Sec. 6). We notice that the number of Lanczos iterations required for convergence is intimately related with the local spectral density. For instance, in the Lanczos procedure with $A = Q$ we see a nice proportionality around $\lambda = 0$.

Apart from the example of Fig. 2, complete spectra were determined on a number of quenched and unquenched configurations with two flavors of dynamical fermions on $4^4$ – $8^3 \cdot 12$ lattices, and almost on a $12^4$ lattice. These configurations are quoted in Tables 4 and 5 below. In nontrivial gauge fields no degeneracy of any eigenvalue was found, neither for $Q$ nor for $Q^2$. In case of the operator $Q$ we always encountered an equal number of negative and positive eigenvalues. Furthermore, no discrepancies were found when the squared eigenvalues of $Q$ were compared with the results for $Q^2$.

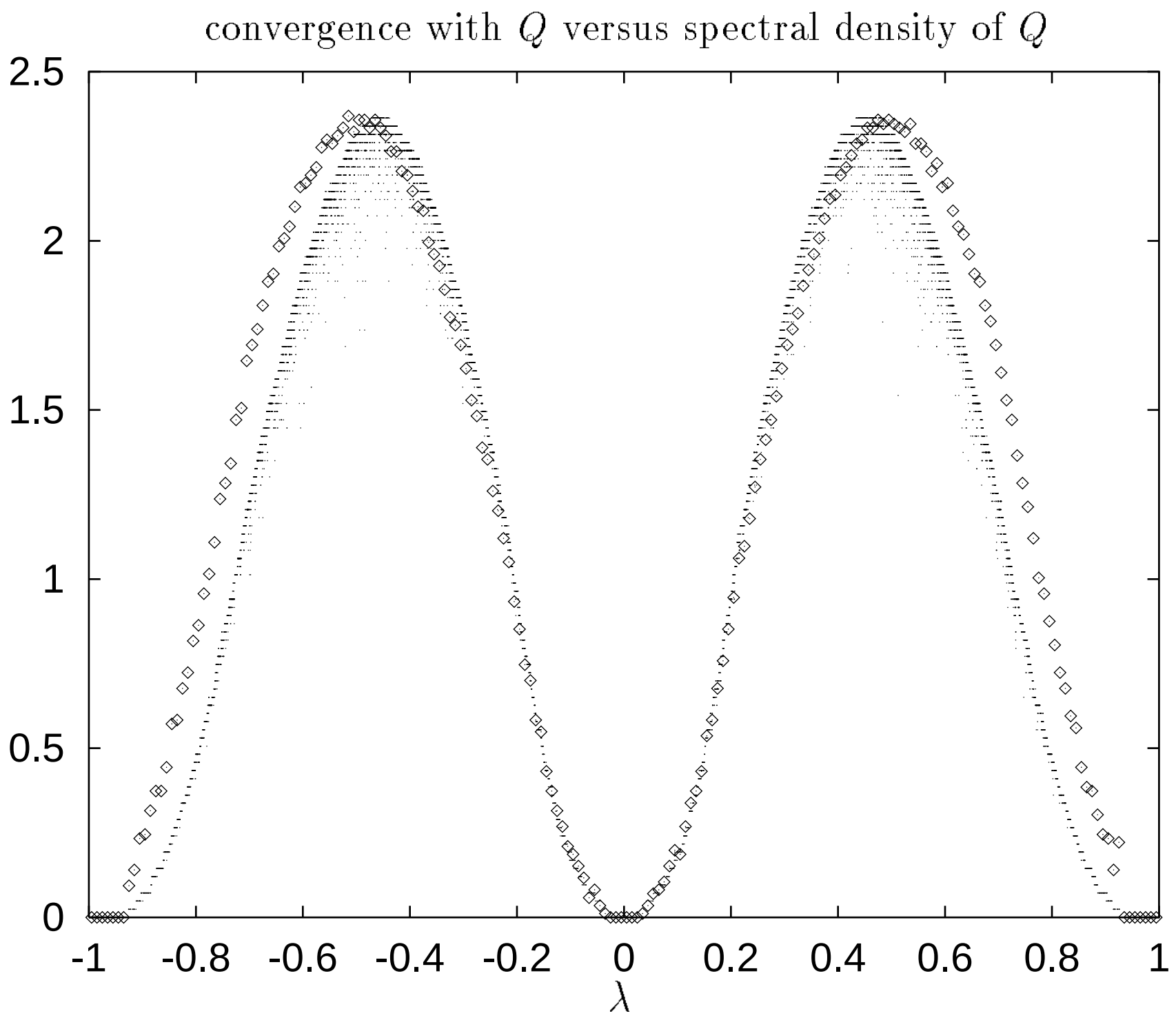

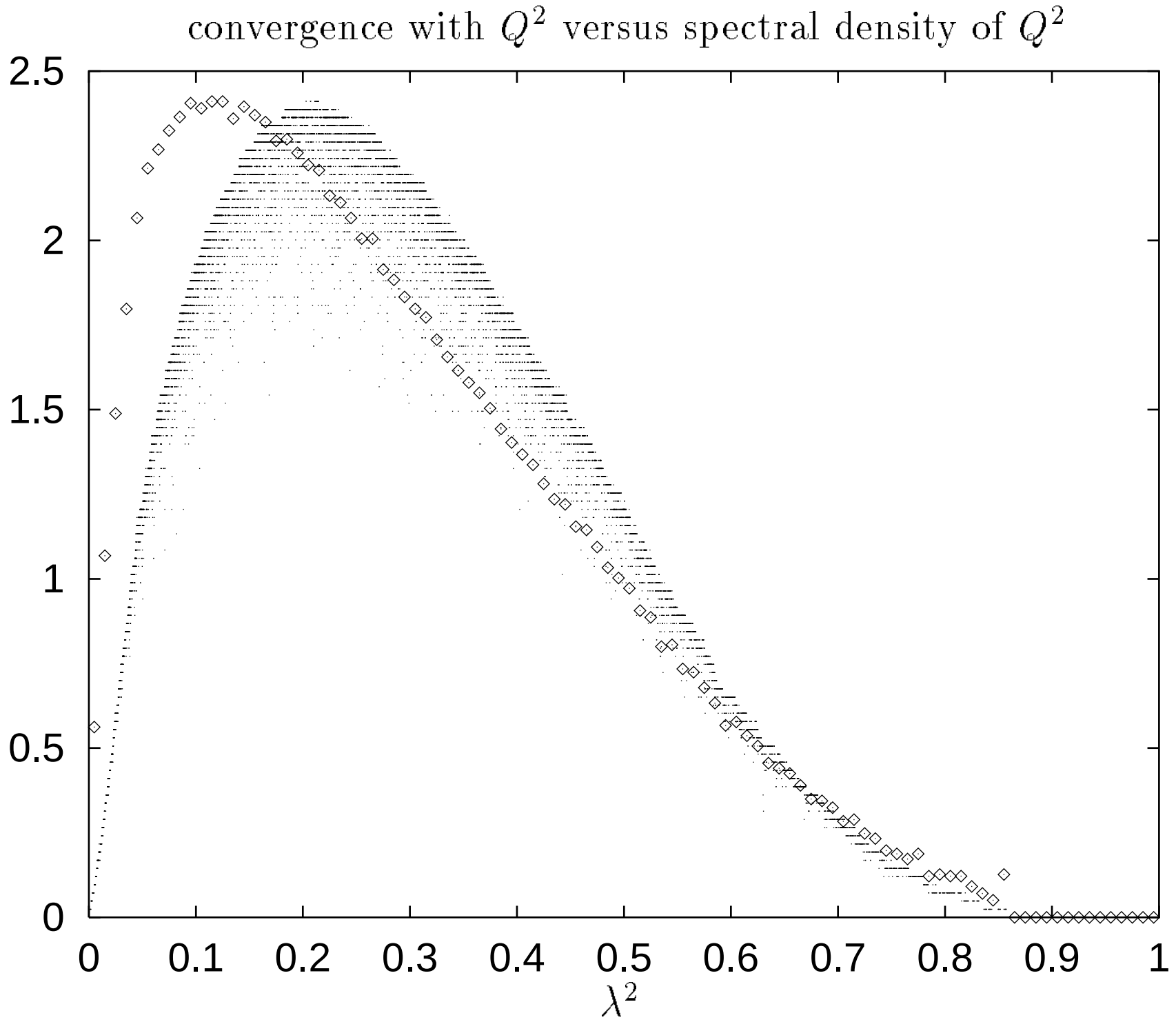

Fig. 2. Convergence of the Lanczos method versus spectral density for $Q$ (top) and for $Q^2$ (bottom) in an unquenched gauge field at $\beta = 2.12$, $\kappa = 0.15$ on a $6^3 \cdot 12$ lattice: Dots give the number $j$ of Lanczos iterations in multiples of $n$ which are required for convergence of $\lambda$ or $\lambda^2$, and diamonds indicate the spectral densities scaled such that the maxima have the same value as the maxima of $j/n$.

Examples for complete spectra are given in Fig. 3. The integrated densities of eigenvalues $N(\lambda)$ and $N(\lambda^2)$ follow directly from the numerical data. They are normalized such that they take values between zero and one. Denoting on a lattice $\Lambda$ of volume $|\Lambda|$ the $k$-th (sorted) eigenvalue of $Q$ by $\lambda_k$, $k = 1, \ldots, n$, $n = 4N_c|\Lambda|$, then $N$ is defined by

$$N(\lambda_k) = k/(4N_c|\Lambda|) \ , \tag{5}$$

and analogously for the squared operator. $N_c$ denotes the number of colours which is two in our case.

We have the following consistency checks which provide good evidence that all computed complete spectra are correct. First, on all investigated $4^4$ – $8^3 \cdot 12$ lattices the correct number of $n$ different eigenvalues was found. Second, we have analytical sum rules for powers of the eigenvalues, cf. also Ref. [18]. The trace of $Q$ equals zero,[6] and the trace of $Q^2$ reads in any unitary gauge field

$$\mathrm{Tr\,Q^2} = 4\,\mathrm{N_c}\,|\Lambda|\,(4 + \frac{1}{4\kappa^2})\,/\,(4 + \frac{1}{2\kappa})^2 \ . \tag{6}$$

Numerically we obtained $|\mathrm{Tr\,Q}| \lesssim 10^{-8}$, and $\mathrm{Tr\,Q^2}$ came out with a relative accuracy of $10^{-8}$ – $10^{-12}$ (decreasing with increasing $|\Lambda|$). One could check for further sum rules by examining traces of higher powers of $Q$. However, except for $\mathrm{Tr\,Q^3} = 0$, compared with an absolute numerical value of $\lesssim 10^{-7}$, such further checks were not performed. As another additional check for correctness one could compare the eigenvalues of $T^{(j)}$ either with those obtained by using a different initial Lanczos vector $v_1$ or with those obtained in a gauge transformed configuration. Such checks were done for just one configuration and it turned out that the converged eigenvalues agree of course, but the spurious eigenvalues are generally dependent on $v_1$ or they are gauge dependent.

We confirmed that the numerical effort for the determination of complete spectra by means of the Lanczos algorithm grows with the square of the lattice volume. In order to obtain all eigenvalues in nontrivial gauge fields it was found empirically that $j = 2\,n$ does not suffice, but $j = 4\,n = 16N_c|\Lambda|$ worked in all cases,[7] both for $Q$ and for $Q^2$. (Actually, the result of Fig. 2 suggests that $j \approx 2.5\,n$ would also work.) With the choice of $j = 4\,n$ the exact eigenvalues close to and furthest from the origin have multiplicities in $T^{(j)}$ of $O(10)$ – $O(100)$. However, the majority of the other correct eigenvalues of $T^{(j)}$ has a multiplicity of only one. Therefore it is generally too costly to accept an eigenvalue of $T^{(j)}$ as correct only when it has replicated. In the general case one better relies on the criterion that an eigenvalue has converged when it is not spurious and does not change with increasing $j$.

[6] Traces are understood over colour and spinor indices.

[7] In case of the squared staggered Dirac operator in SU(2) gauge fields, where every eigenvalue has a multiplicity of four, one obtains the complete spectrum when $j$ equals only twice the number of different eigenvalues, i.e. when $j = |\Lambda|$ [18].

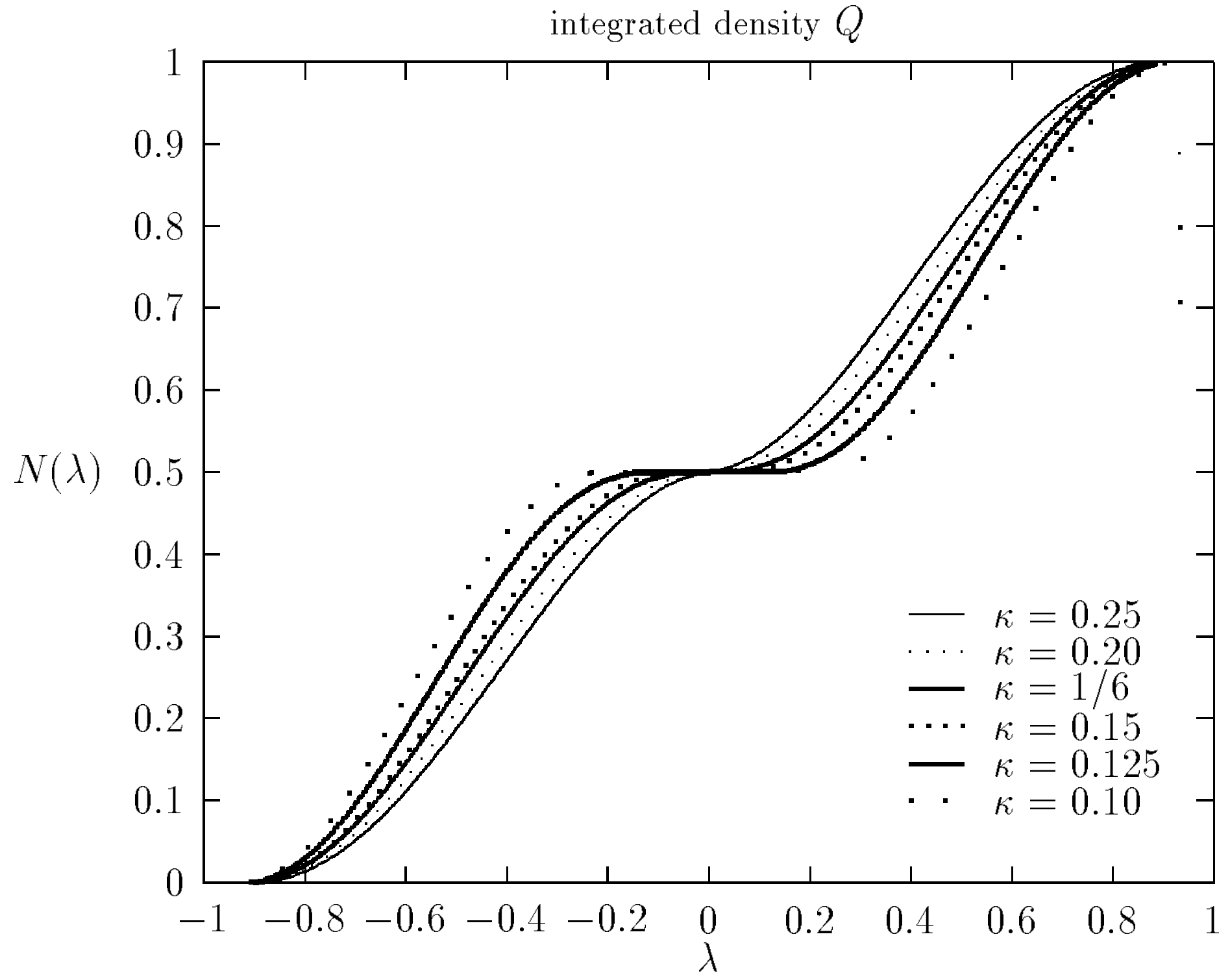

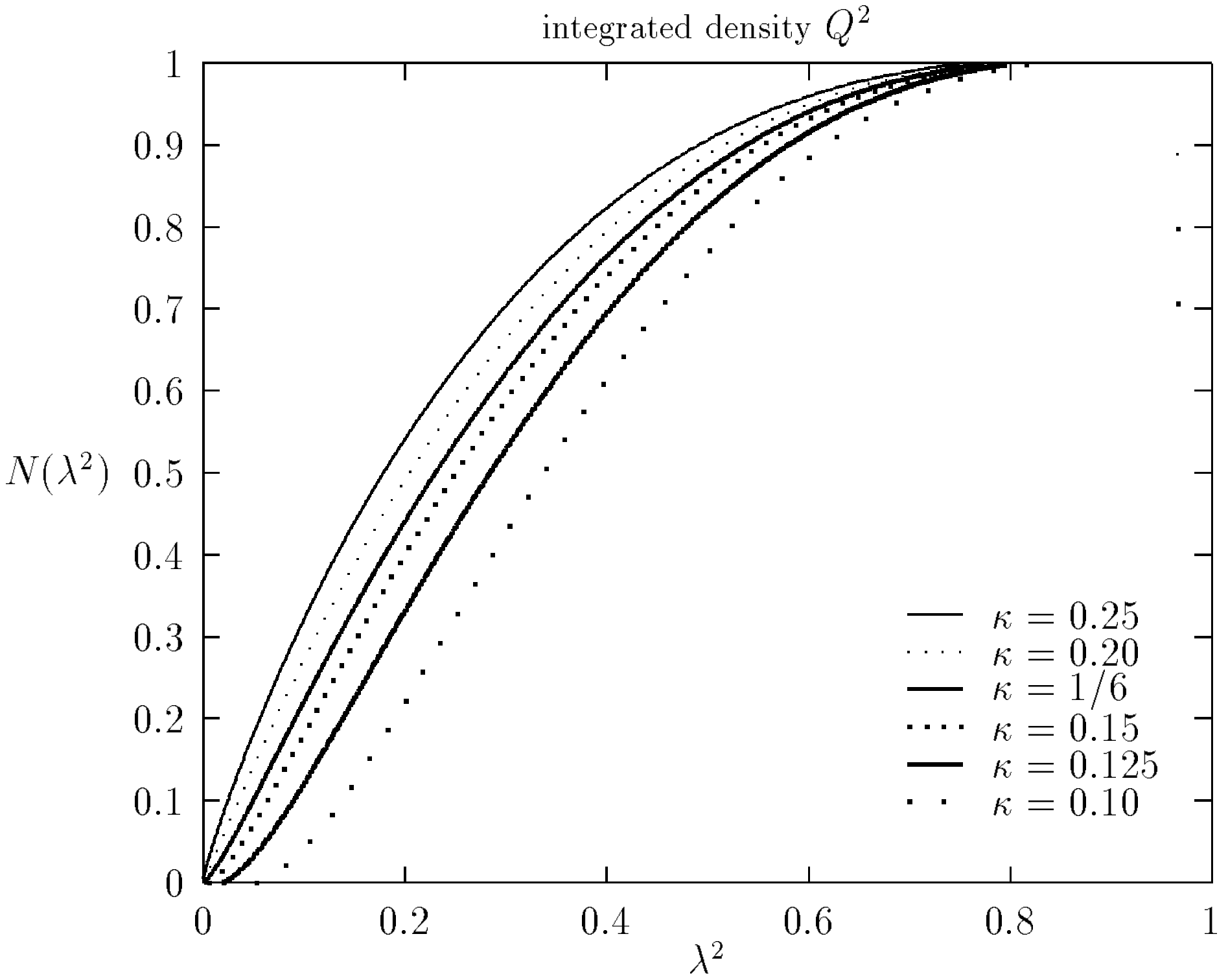

Fig. 3. Integrated spectral densities of $Q$ (top) and $Q^2$ (bottom). The results are obtained in a quenched gauge field at $\beta = 1.8$ on a $6^4$ lattice. At $\kappa = 0.15$ curves of a quenched $12^4$ gauge field at $\beta = 1.8$ and unquenched gauge fields at $\beta = 2.12$ on $6^3 \cdot 12$ and $8^3 \cdot 12$ lattices coincide (on the scale of the figure) with the result on the $6^4$ lattice.

Finally, we mentioned that the complete spectrum could almost be determined on a $12^4$ lattice. Because of memory limitations the largest matrix $T^{(j)}$ which could be handled is restricted to $j = 21 \cdot 12^4 = 2.625\,n$. With this choice of $j$ we investigated a quenched $12^4$ gauge field at $\beta = 1.8$, $\kappa = 0.15$. The computations required something like two weeks of workstation CPU time. (Six hours are required for the 100 lowest eigenvalues of $Q^2$, cf. Table 1.) All eigenvalues of $Q$ but four could be found, i.e. a total of 165884 out of 165888. For the traces these 165884 values yielded $\sum \lambda_k = -0.09087$ and $\sum \lambda_k^3 = -0.04237$ instead of zero, and $\sum \lambda_k^2 = 46612.04$ instead of 46613.15. When one computes with these 165884 values the normalized integrated densities $N(\lambda)$ and $N(\lambda^2)$, one encounters no difference compared with the result on a $6^4$ lattice, only statistical noise which can be recognized on the $6^4$ lattice is completely smoothed out on the $12^4$ lattice.

## 5 Fermionic determinants

In this section we give results for fermionic determinants of the configurations where all eigenvalues were computed. Besides the results for Wilson fermions we also quote values of determinants of staggered fermions whose spectra were determined in a previous publication [18]. It must be stressed that all numbers listed here are only examples for individual configurations.

### *5.1 Wilson fermions*

We start by referring to free Wilson determinants in Table 3. In this case the eigenvalues of $Q$ are known analytically,

$$\lambda = \pm \left[ \left(\frac{1}{2\kappa} - \sum_{\mu=1}^{4} \cos p_\mu\right)^2 + \sum_{\mu=1}^{4} \sin^2 p_\mu \right]^{1/2} \Big/ \left(4 + \frac{1}{2\kappa}\right) \qquad (7)$$

where each $\lambda$ appears twice with either sign, and $p_\mu = 2\pi k_\mu / L_\mu$ with $k_\mu = 0, 1, \ldots, L_\mu - 1$ and $L_\mu$ is the extension of the lattice in $\mu$-direction.

Table 3
Free Wilson fermions (in a trivial gauge field): $(\log_{10}\det Q)/(4N_c|\Lambda|)$. In case of $\kappa = 0.125$ zero modes are excluded from the determinant.

| $\|\Lambda\|$ | $\kappa = 0.1$ | $\kappa = 0.125$ | $\kappa = 0.15$ | $\kappa = 1/6$ | $\kappa = 0.20$ | $\kappa = 0.25$ |
|---|---|---|---|---|---|---|
| $4^4$ | -0.254927 | -0.293846 | -0.332655 | -0.350106 | -0.376619 | -0.639097 |
| $6^3 \cdot 12$ | -0.253839 | -0.295955 | -0.329555 | -0.346555 | -0.371246 | -0.415416 |
| $8^4$ | -0.253802 | -0.295996 | -0.329410 | -0.346378 | -0.370939 | -0.406257 |
| $12^4$ | -0.253797 | -0.296251 | -0.329390 | -0.346351 | -0.370860 | -0.394140 |
| $16^4$ | -0.253797 | -0.296301 | -0.329390 | -0.346350 | -0.370856 | -0.392127 |
| $32^4$ | -0.253797 | -0.296325 | -0.329390 | -0.346350 | -0.370856 | -0.391263 |

Table 4
Examples for Wilson fermions in quenched SU(2) gauge fields:
$(\log_{10}\det Q)/(4N_c|\Lambda|)$. The case $\beta = 0$ corresponds to a random gauge field.

| $\|\Lambda\|$ | $\beta$ | $\kappa = 0.1$ | $\kappa = 0.125$ | $\kappa = 0.15$ | $\kappa = 1/6$ | $\kappa = 0.20$ | $\kappa = 0.25$ |
|---|---|---|---|---|---|---|---|
| $6^4$ | 1.80 | -0.254757 | -0.299665 | -0.339289 | -0.362812 | -0.402336 | -0.440477 |
| $6^4$ | 2.80 | -0.254331 | -0.298278 | -0.334999 | -0.354646 | -0.383828 | -0.409753 |
| $6^4$ | 0.00 | -0.255279 | -0.301046 | -0.342455 | -0.368026 | -0.415065 | -0.477151 |
| $8^4$ | 0.00 | -0.255277 | -0.301041 | -0.342445 | -0.368010 | -0.415038 | -0.476955 |

Quenched results are contained in Table 4, while unquenched results are collected in Table 5. One concludes a nice exponential dependence of the determinant on the lattice volume, already for relatively small lattices.

In a quenched Monte Carlo simulation the fermionic determinant is kept at a fixed value. From Table 5 one can get a feeling for the fluctuations of the determinant in an unquenched run. We can compare the values of eight independent dynamical $6^3 \cdot 12$ configurations at $\beta = 2.12$, $\kappa = 0.15$. The logarithmic entries in Table 5 fluctuate by $\approx 0.0003$ which translates to a fluctuation of the determinant itself by six orders of magnitude on the $6^3 \cdot 12$ lattice.

Table 5
Examples for Wilson fermions in unquenched SU(2) gauge fields with two flavours of dynamical fermions: $(\log_{10} \det Q)/(4N_c|\Lambda|)$.

| $\|\Lambda\|$ | $\beta$ | $\kappa = 0.15$ |
|---|---|---|
| $4^4$ | 1.75 | -0.338947 |
| $6^3 \cdot 12$ | 2.12 | -0.337169 |
| $6^3 \cdot 12$ | 2.12 | -0.337129 |
| $6^3 \cdot 12$ | 2.12 | -0.337282 |
| $6^3 \cdot 12$ | 2.12 | -0.337388 |
| $6^3 \cdot 12$ | 2.12 | -0.337385 |
| $6^3 \cdot 12$ | 2.12 | -0.337266 |
| $6^3 \cdot 12$ | 2.12 | -0.337269 |
| $6^3 \cdot 12$ | 2.12 | -0.337235 |
| $8^3 \cdot 12$ | 2.12 | -0.337295 |

Table 6
Free staggered fermions (in a trivial gauge field): $[\log_{10} \det(-\not{D}^2_{\rm stag} + m^2)]/(N_c|\Lambda|)$. In case of $m = 0$ zero modes are excluded from the determinant.

| $\|\Lambda\|$ | $m = 0$ | $m = 0.05$ | $m = 0.2$ |
|---|---|---|---|
| $6^4$ | 0.878691 | 0.835873 | 0.853054 |
| $12^4$ | 0.869726 | 0.867211 | 0.870592 |
| $24^4$ | 0.868572 | 0.868572 | 0.871123 |
| $32^4$ | 0.868501 | 0.868615 | 0.871133 |
| $48^4$ | 0.868471 | 0.868628 | 0.871135 |
| $64^4$ | 0.868465 | 0.868630 | 0.871135 |
| $80^4$ | 0.868463 | 0.868630 | 0.871135 |

Table 7
Examples for staggered fermions in nontrivial SU(2) gauge fields: $[\log_{10}\det(-\not{D}_{\rm stag}^2 + m^2)]/(N_{\rm c}|\Lambda|)$. The case $\beta = 0$ corresponds to a random gauge field.

| $\|\Lambda\|$ | $\beta$ | $m = 0$ | $m = 0.05$ | $m = 0.2$ |
|---|---|---|---|---|
| quenched | | | | |
| $6^4$ | 0.0 | 0.498590 | 0.511586 | 0.553658 |
| $6^4$ | 1.8 | 0.639018 | 0.645632 | 0.670574 |
| $6^4$ | 2.8 | 0.785223 | 0.785556 | 0.790327 |
| $12^4$ | 0.0 | 0.499169 | 0.513180 | 0.554995 |
| $12^4$ | 1.8 | 0.640427 | 0.648199 | 0.672902 |
| $12^4$ | 2.0 | 0.674723 | 0.680302 | 0.699873 |
| $12^4$ | 2.4 | 0.757622 | 0.758471 | 0.765383 |
| $12^4$ | 2.6 | 0.777512 | 0.777890 | 0.782955 |
| $12^4$ | 2.7 | 0.784554 | 0.784866 | 0.789400 |
| $12^4$ | 2.8 | 0.788623 | 0.788923 | 0.793297 |
| unquenched | | | | |
| $6^4$ | 1.8 | – | 0.697371 | 0.704457 |
| $6^4$ | 2.8 | – | 0.799398 | 0.802219 |
| $12^4$ | 1.8 | – | 0.638295 | 0.707586 |
| $12^4$ | 2.4 | – | 0.717424 | 0.776017 |
| $12^4$ | 2.8 | – | 0.779199 | 0.798827 |

### *5.2 Staggered fermions*

The Lanczos algorithm was studied earlier for the staggered Dirac operator $\not{D}_{\rm stag}$ (also with periodic boundary conditions) in Ref. [18], but no determinants were computed there. Here we use the old data and for a sake of completeness quote results in Tables 6 and 7. Note that the staggered quark mass $m$ is measured in units of twice the separation of neighbouring sites,[8] i.e. in a more frequent convention one would quote $m = 0$, 0.025, 0.01 instead of 0, 0.05, 0.02, respectively.

[8] The reason is that free staggered fermions enjoy translation invariance only by shifts of two lattice spacings [20].

We add as a remark that the staggered data of Ref. [18] were further analysed by Halasz and Verbaarschot in [8]. These authors showed that the eigenvalue correlations of the staggered Dirac operator in SU(2) gauge fields are very well described by the Gaussian symplectic ensemble of a random matrix model with the chiral symmetry of the Dirac operator built in.

## 6 Spectral densities

The density of eigenvalues of $Q^2$ around zero can be related with the chiral limit, if one connects the spectral density of $Q^2$ with that of the Dirac operator (not multiplied by $\gamma_5$) in the spirit of the "pion norm" of Ref. [1]. Such an analysis will however be done elsewhere. From the algorithmic point of view spectral densities are interesting because they determine convergence properties, as we saw from Fig. 2.

Figs. 4 – 6 show results for normalized spectral densities as obtained in individual gauge field configurations of the present algorithmic investigation. The data points in these figures were obtained by a simple binning procedure, i.e. approximating $\rho(\lambda) \equiv \mathrm{dN}(\lambda)/\mathrm{d}\lambda$ by $\triangle N(\lambda)/\triangle\lambda$ with $\triangle\lambda = 0.01$, and analogously for the spectral density $\rho(\lambda^2)$ of $Q^2$. The solid curves indicate everywhere reference data of unquenched gauge fields at $\kappa = 0.15$, $\beta = 2.12$ on $6^3 \cdot 12$ and $8^3 \cdot 12$ lattices, where numerical results coincide.

The information in each of Figs. 4 – 6 is redundant because of the relation $\rho(\lambda^2) = \rho(\lambda)/(2\lambda)$, which can be verified numerically to a very high precision. Despite this redundancy we consider it worthwhile to present results for $\rho(\lambda)$ as well as for $\rho(\lambda^2)$. Note for instance that in the $\kappa$-range investigated $\rho(\lambda^2)$ in Fig. 4 stays finite as $\lambda^2 \to 0$ because $\rho(\lambda = 0) = 0$. On the other hand, $\rho(\lambda^2)$ must diverge for $\lambda^2 \to 0$ if $\rho(\lambda = 0)$ is finite. This is the case for $\kappa = 0.25$ in Figs. 5 and 6, and the case $\kappa = 0.20$ in Fig. 5 seems to be close to a marginal situation.

Remarkably, wherever we are able to compare data from different lattice sizes $\geq 6^4$ which correspond to the same physical situation, we cannot find finite size effects in the normalized spectral densities (using $\triangle\lambda = 0.01$), and also in the integrated densities on the scale of Fig. 3; the only effect of a larger lattice is to smooth out statistical fluctuations. This is true for the unquenched gauge fields at $\kappa = 0.15$, $\beta = 2.12$ on $6^3 \cdot 12$ and $8^3 \cdot 12$ lattices, for quenched gauge fields at $\beta = 1.8$, $\kappa = 0.15$ on $6^4$ and $12^4$ lattices, and for random gauge fields ($\beta = 0$) on $6^4$ and $8^4$ lattices in the $\kappa$-range $0.10, \ldots, 0.25$.

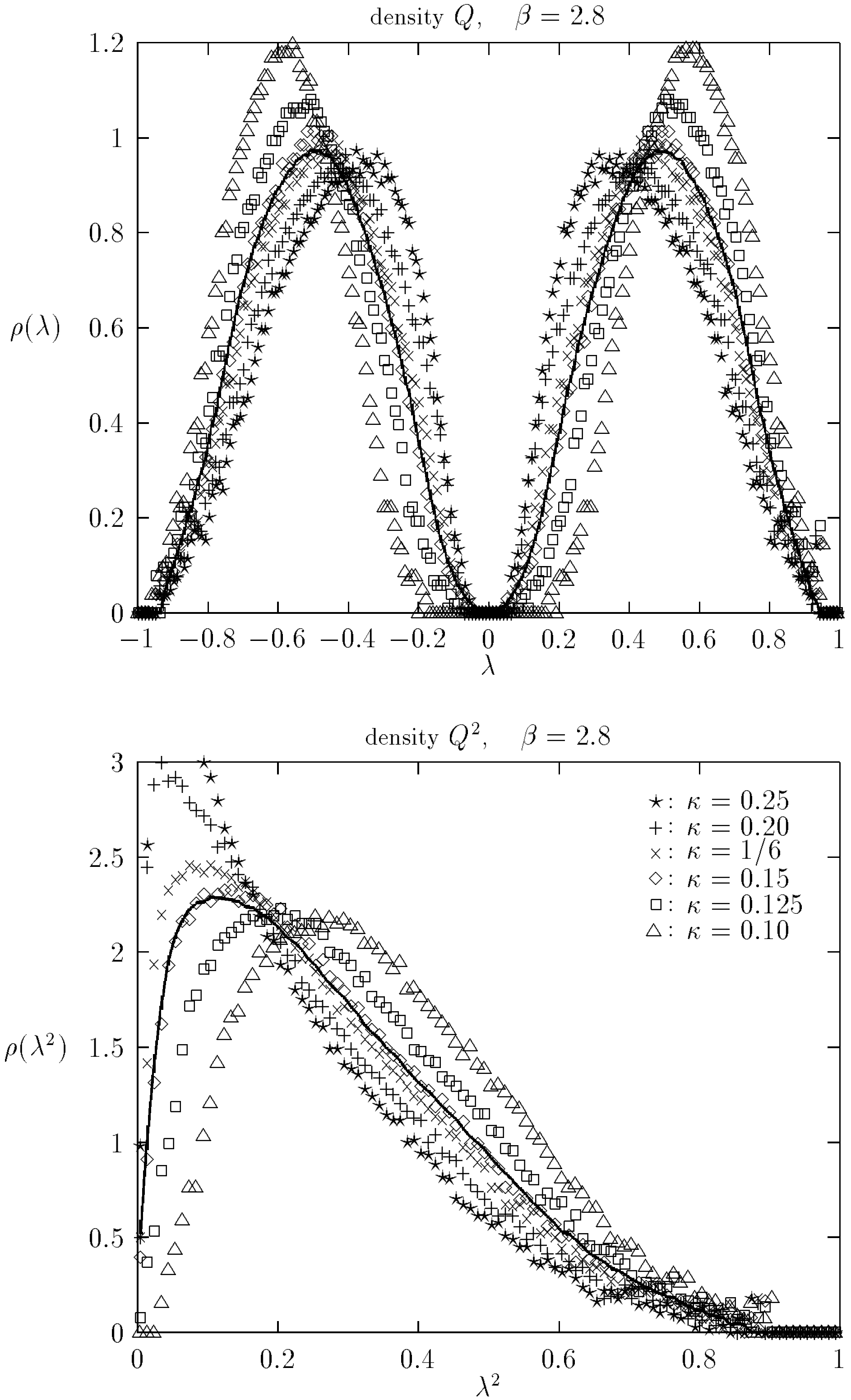

Fig. 4. Spectral density of $Q$ (top) and of $Q^2$ (bottom) as obtained in a quenched gauge field at $\beta = 2.8$ on a $6^4$ lattice. The solid lines indicate the densities in unquenched gauge fields at $\kappa = 0.15$, $\beta = 2.12$ on $6^3 \cdot 12$ and $8^3 \cdot 12$ lattices.

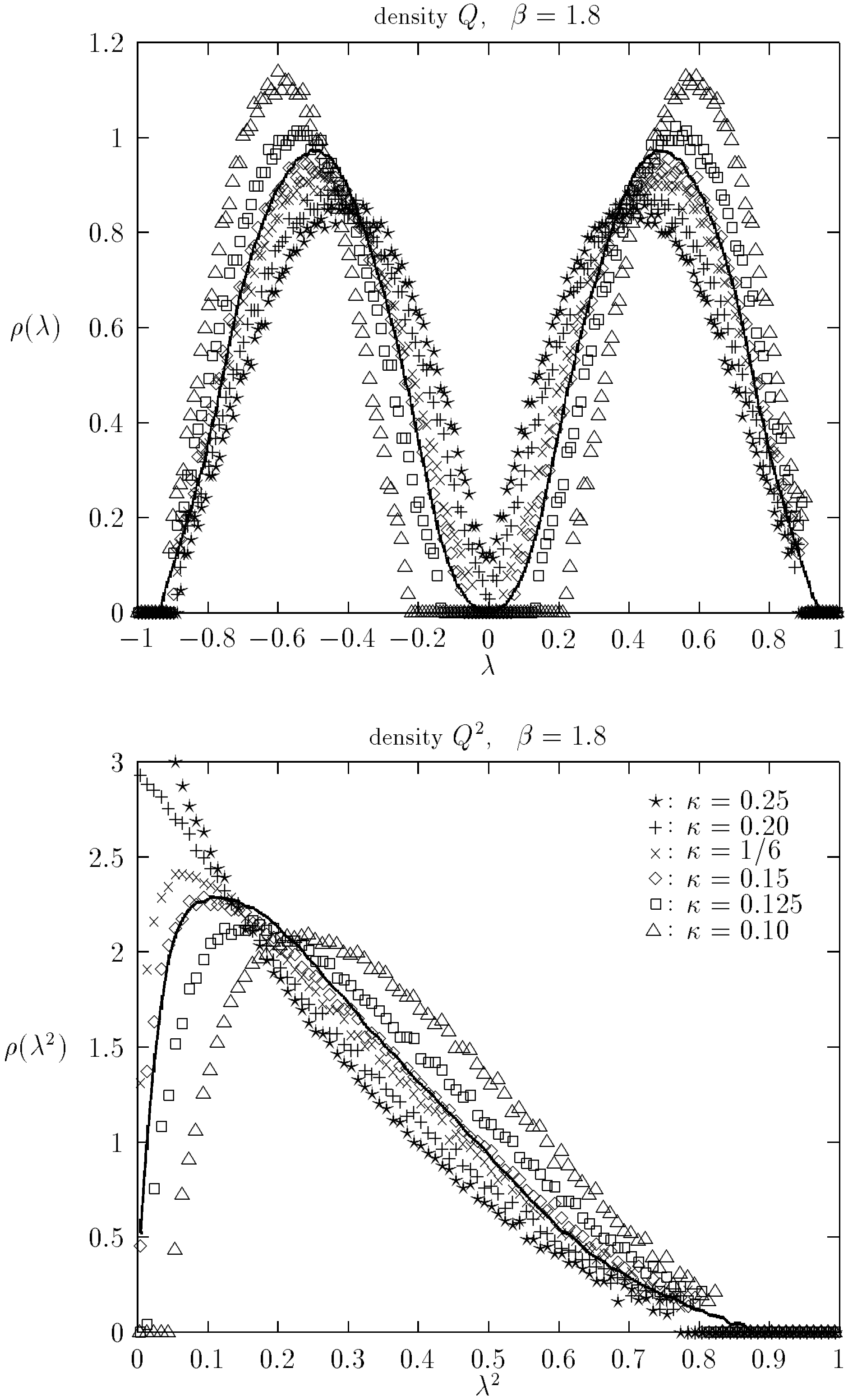

Fig. 5. Spectral density of $Q$ (top) and of $Q^2$ (bottom) as obtained in a quenched gauge field at $\beta = 1.8$ on a $6^4$ lattice. The curves at $\kappa = 0.15$ coincide with the results on a quenched $12^4$ lattice at $\beta = 1.8$ (except that fluctuations are smoothed out). The solid lines indicate the densities in unquenched gauge fields at $\kappa = 0.15$, $\beta = 2.12$ on $6^3 \cdot 12$ and $8^3 \cdot 12$ lattices.

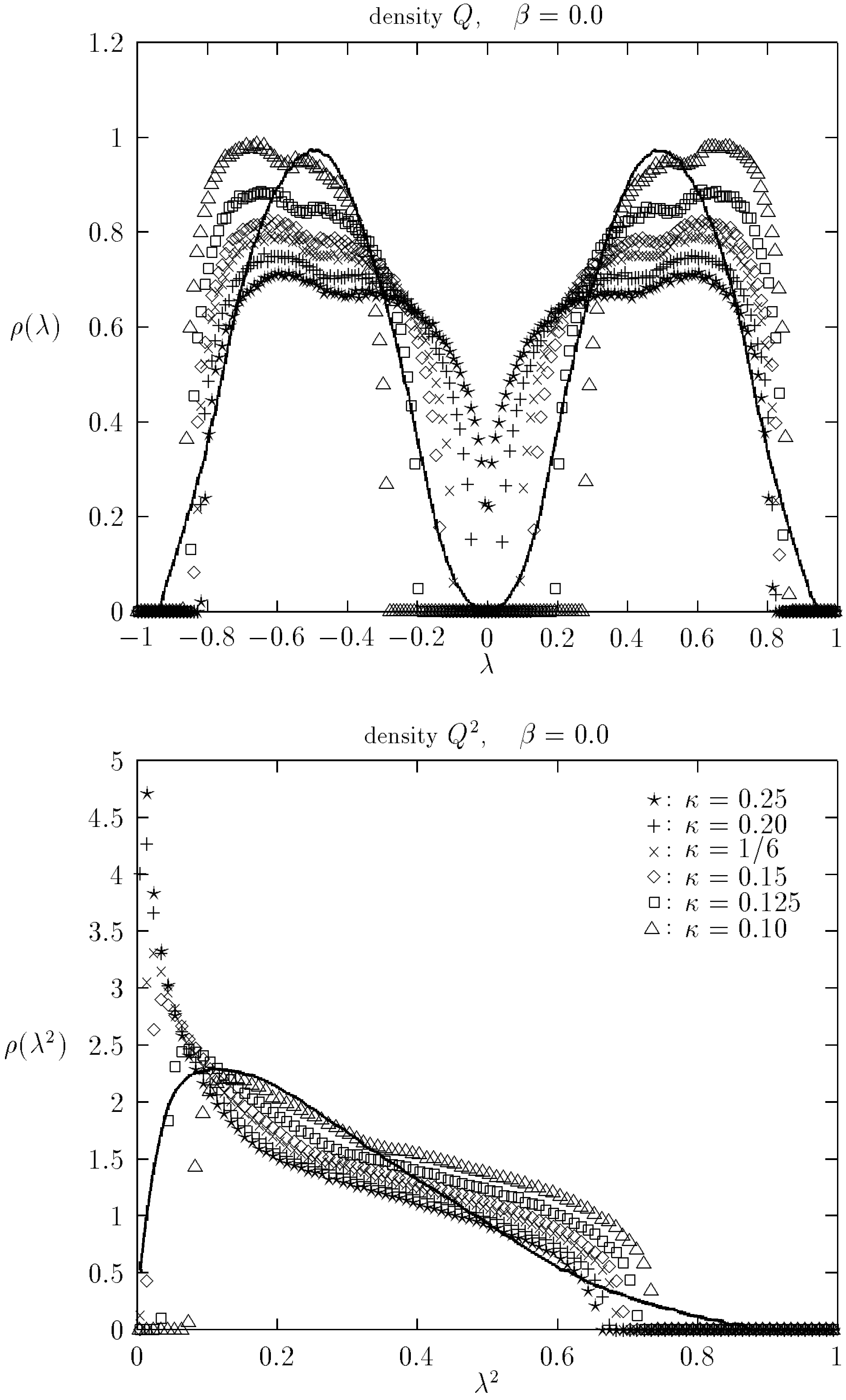

Fig. 6. Spectral density of $Q$ (top) and of $Q^2$ (bottom; note the different scale of the ordinate compared with Figs. 4 and 5) as obtained in a random gauge field ($\beta = 0.0$). There is no distinction between $6^4$ and $8^4$ lattices, except that fluctuations are smoothed out on the larger lattice. The solid lines indicate the densities in unquenched gauge fields at $\kappa = 0.15$, $\beta = 2.12$ on $6^3 \cdot 12$ and $8^3 \cdot 12$ lattices.

Finally, let us resume the discussion about the relation between convergence and the spectral density. We noted that the complete spectrum could be obtained for any of the configurations studied if $j = 4\,n$, both for $Q$ and for $Q^2$. This means that a complete computation is not affected by the degree of criticality of the system, because we included random gauge fields with $\kappa = 0.25$ which is believed to be a critical point (in strong coupling and in the large $N_c$ limit) [21]. We also studied random configurations with $\kappa$ in the neighbourhood of 0.25 and up to 2.0, and we did not encounter any exceptional example where not all eigenvalues were detected with $j = 4\,n$. However, the picture changes if we pay attention only to some of the low-lying eigenvalues of $Q^2$. In this case one can witness quite a drastic difference between the Lanczos procedure with $A = Q$ and $A = Q^2$ in the vicinity of a critical point. It takes much longer until the first few eigenvalues have converged when one runs the Lanczos method with $Q^2$ instead of $Q$. The reason is obvious when one looks at the spectral densities near zero in Fig. 6 and recalls Fig. 2.

## 7 Conclusions and comparison with the algorithm of Ref. [16]

If one is willing to pay the necessary CPU time — which increases from 2.5 workstation minutes on a $4^4$ lattice to more than two workstation weeks on a $12^4$ lattice — one can compute the complete spectrum of $Q$ in quenched and unquenched SU(2) gauge fields by means of Cullum's and Willoughby's Lanczos procedure. Such a computation of the complete spectrum is also reliable in cases where the configuration is in the immediate neighbourhood of a critical point, as was seen in random gauge fields around $\kappa = 0.25$. We always detected the correct number of eigenvalues which equals the dimension of the Dirac operator (except for four missing eigenvalues on the $12^4$ lattice, but this was due to memory limitations), and we were able to check the correctness of these eigenvalues by analytical sum rules.

With data for complete spectra one can compute fermionic determinants and spectral densities. For the determinants we found a nice exponential dependence on the lattice volume already for relatively small lattices. The spectral densities are expected to contain information about the question of chiral symmetry breaking and the phase structure of the theory. Possibly, with data for all eigenvalues one can also investigate questions of universality similarly to Ref. [8]. However, such physical applications will be addressed elsewhere.

In Fig. 2 we showed that the convergence behaviour of the Lanczos method is intimately linked with the local spectral density $\rho$ over the entire spectrum. Convergence is fastest in those regions of the spectrum where the density of eigenvalues is lowest. Thus, if one is interested in the low-lying eigenvalues of $Q^2$ one should run the Lanczos procedure with $Q$ and not with $Q^2$, because of the relation $\rho(\lambda^2) = \rho(\lambda)/(2\lambda)$. The difference in performance is roughly a factor of two in the computational cost (measured in the number of $Q\times$vector multiplications) when the system studied is not very close to criticality, otherwise the cost ratio gets even bigger.

Finally, let us turn to a comparison of the Lanczos procedure with the accelerated conjugate gradient algorithm of Ref. [16] for the computation of low-lying eigenvalues of $Q^2$. In Ref. [16] the same unquenched configurations as in the present paper were studied and a number of low-lying eigenvalues with a relative accuracy of $10^{-4}$ were computed. We find the following results. First, by means of the reference data of [16] we are able to confirm that in the tests performed the Lanczos procedure with a small number of iterations $j \ll n$ never skipped an eigenvalue with small modulus. The reason is that in the regions of lowest spectral density the Lanczos method shows an "ordered" convergence by which we mean that between any two converged eigenvalues there is no eigenvalue which has not yet converged also or which will not converge within a few more iterations. Second, comparing the performance measured in $Q\times$vector multiplications (which require the major fraction of the total CPU time), the accelerated conjugate gradient algorithm is inferior to the Lanczos method by about a factor of $5-8$. Possibly, this factor might be larger for systems close to criticality where the spectral density favours computations with $Q$ instead of $Q^2$. Hence, a practitioner may prefer the Lanczos method provided he has a priori information about degeneracies in the spectrum. However, from a rigorous point of view the conjugate gradient approach is favourable because it yields all eigenvalues with a rigorous straightforward and practical error bound, and all eigenvalues are detected with their correct multiplicities. Furthermore, approximations to eigenvectors are obtained as a by-product. This latter point is a clear advantage for applications where not only eigenvalues but also eigenvectors are needed, e.g. when one is interested in the contribution of the low-lying eigenmodes to physical observables.

## Acknowledgement

I wish to thank B. Bunk, K. Jansen, H. Simma, and U. Wolff for discussions, and P. Weisz for confirming the free Wilson determinants. C. Liu and K. Jansen are thanked for providing the unquenched Wilson configurations. Financial support by Deutsche Forschungsgemeinschaft under grant Wo 389/3-1 is gratefully acknowledged. The computations reported here were performed on HP workstations of the Humboldt-University.

## References

[1] K.M. Bitar, A.D. Kennedy, and P. Rossi, *The chiral limit and phase structure of QCD with Wilson fermions*, Phys. Lett. B234 (1990) 333.

[2] S.J. Hands and M. Teper, *On the value and origin of the chiral condensate in quenched* SU(2) *lattice gauge theory*, Nucl. Phys. B347 (1990) 819.

[3] I. Barbour, E. Laermann, T. Lippert, and K. Schilling, *Towards the chiral limit with dynamical blocked Wilson fermions*, Phys. Rev. D46 (1992) 3618.

[4] H. Leutwyler and A. Smilga, *Spectrum of Dirac operator and role of winding number in QCD*, Phys. Rev. D46 (1992) 5607;

A.V. Smilga, *Chiral symmetry and spectrum of euclidean Dirac operator in QCD*, hep-th/9503049 (March 1995), contributed to International Workshop on Nuclear and Particle Physics: Chiral Dynamics in Hadrons and Nuclei, Seoul, Korea, 6-10 Feb 1995.

[5] A. Hoferichter, V.K. Mitrjushkin, and M. Müller-Preussker, *On the chiral limit in lattice gauge theories with Wilson fermions*, hep-lat/9506006, and preprint HUB-IEP-95/5 (May 1995).

[6] F.X. Lee, H.D.Trottier, and R.M. Woloshyn, *Abelian dominance of chiral symmetry breaking in lattice QCD*, hep-lat/9509028.

[7] J.J.M. Verbaarschot, *Chiral random matrix theory and QCD*, hep-th/9405006 (May 1994).

[8] M.A. Halasz and J.J.M. Verbaarschot, *Universal fluctuations in spectra of the lattice Dirac operator*, Phys. Rev. Lett. 74 (1995) 3920.

[9] M. Lüscher, *A new approach to the problem of dynamical quarks in numerical simulations of lattice QCD*, Nucl. Phys. B418 (1994) 637.

[10] B. Bunk, K. Jansen, B. Jegerlehner, M. Lüscher, H. Simma, and R. Sommer, *A new simulation algorithm for lattice QCD with dynamical quarks*, Nucl. Phys. B (Proc. Suppl.) 42 (1995) 49.

[11] C. Alexandrou, A. Borelli, Ph. de Forcrand, A. Galli, and F. Jegerlehner, *Full QCD with the Lüscher local bosonic action*, hep-lat/9506001 (June 1995).

[12] G.H. Golub and C.F. v. Loan, *Matrix computations*, second edition, (The Johns Hopkins University Press, Baltimore, 1990).

[13] I.M. Barbour, N.-E. Behilil, P.E. Gibbs, G. Schierholz, and M. Teper, *The Lanczos method in lattice gauge theories*, in: The recursion method and its applications, Springer Series in Solid-State Sciences 58, eds. D. G. Pettifor and D. L. Weaire (Springer-Verlag, Berlin, 1985).

[14] R. Setoodeh, C.T.H. Davies, and I.M. Barbour, *Wilson fermions on the lattice – a study of the eigenvalue spectrum*, Phys. Lett. B213 (1988) 195.

[15] B. Bunk, K. Jansen, M. Lüscher, and H. Simma, *Conjugate gradient algorithm to compute the low-lying eigenvalues of the Dirac operator in lattice QCD*, internal DESY-report (September 1994).

[16] T. Kalkreuter and H. Simma, *An accelerated conjugate gradient algorithm to compute low-lying eigenvalues – a study for the Dirac operator in* SU(2) *lattice QCD*, hep-lat/9507023, and preprints DESY 95-137, HUB-IEP-95/10 (July 1995).

[17] J. Cullum and R.A. Willoughby, *Computing eigenvalues of very large symmetric matrices – an implementation of a Lanczos algorithm with no reorthogonalization*, J. Comp. Phys. 44 (1981) 329.

[18] T. Kalkreuter, *Spectrum of the Dirac operator and multigrid algorithm with dynamical staggered fermions*, Phys. Rev. D51 (1995) 1305;

*Spectrum of the Dirac operator and inversion algorithms with dynamical staggered fermions*, Nucl. Phys. B (Proc. Suppl.) 42 (1995) 882.

[19] W.H. Press, S.A. Teukolsky, W. T. Vetterling, and B.P. Flannery, *Numerical recipes*, 2nd ed. (Cambridge University Press, Cambridge, 1992).

[20] H. Joos and M. Schaefer, *The representation theory of the symmetry group of lattice fermions as a basis for kinematics in lattice QCD*, Z. Phys. C34 (1987) 465.

[21] N. Kawamoto, *Towards the phase structure of euclidean lattice gauge theories with fermions*, Nucl. Phys. B190 [FS3] (1981) 617.